# Negative Refraction Makes a Perfect Lens

*JB Pendry*
*The Blackett Laboratory*
*Imperial College*
*London SW7 2BZ*
*UK*

*Abstract*

With a conventional lens sharpness of the image is always limited by the wavelength of light. An unconventional alternative to a lens, a slab of negative refractive index material, has the power to focus all Fourier components of a 2D image, even those that do not propagate in a radiative manner. Such super lenses can be realised in the microwave band with current technology and a version operating at the frequency of visible light, but at short distances of a few nanometres, can be realised in the form of a thin slab of silver as our simulations show.

Optical lenses have for centuries been one of scientists' prime tools. Their operation is well understood on the basis of classical optics: curved surfaces focus light by virtue of the refractive index contrast. Equally their limitations are dictated by wave optics: no lens can focus light onto an area smaller than a square wavelength. What is there new to say other than to polish the lens more perfectly and to invent slightly better dielectrics? In this letter I want to challenge the traditional limitation on lens performance and propose a class of super lenses, and to suggest a practical scheme for implementing such a lens.

First let us look more closely at the reasons for limitation in performance. Consider an infinitesimal dipole of frequency $\omega$ in front of a lens. The electric component of the field will be given by some 2D Fourier expansion,

$$\mathbf{E}(\mathbf{r},t) = \sum_{\sigma, k_x, k_y} \mathbf{E}_\sigma (k_x, k_y) \exp(ik_z z + ik_x x + ik_y y - i\omega t) \tag{1}$$

where we choose the axis of the lens to be the $z$-axis. Maxwell's equations tell us that,

$$k_z = +\sqrt{\omega^2 c^{-2} - k_x^2 - k_y^2}, \quad \omega^2 c^{-2} > k_x^2 + k_y^2, \tag{2}$$

The function of the lens is to apply a phase correction to each of the Fourier components so that at some distance beyond the lens the fields reassemble to a focus and an image of the dipole source appears. Except that something is missing: for larger values of the transverse wave vector,

$$k_z = +i\sqrt{k_x^2 + k_y^2 - \omega^2 c^{-2}}, \quad \omega^2 c^{-2} < k_x^2 + k_y^2. \tag{3}$$

These evanescent waves decay exponentially with $z$ and no phase correction will restore them to their proper amplitude. They are effectively removed from the image which generally comprises only the propagating waves. Since the propagating waves are limited to,

$$k_x^2 + k_y^2 < \omega^2 c^{-2} \tag{4}$$

the maximum resolution in the image can never be greater than,

$$\Delta \approx \frac{2\pi}{k_{\max}} = \frac{2\pi c}{\omega} = \lambda \tag{5}$$

and this is true however perfect the lens and however large the aperture.

There is an unconventional alternative to a lens. Material with negative refractive index will focus light even when in the form of a parallel sided slab of material. In figure 1 I sketch the focusing action of such a slab, assuming that the refractive index,

$$n = -1 \tag{6}$$

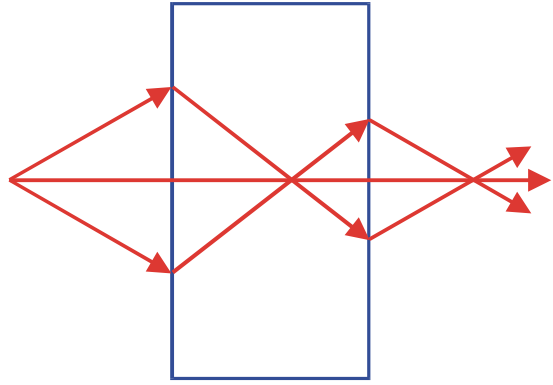

Figure 1. A negative refractive index medium bends light to a negative angle with the surface normal. Light formerly diverging from a point source is set in reverse and converges back to a point. Released from the medium the light reaches a focus for a second time.

A moments thought will show that the figure obeys Snell's law for refraction at the surface as light inside the medium makes a negative angle with the surface normal. The other characteristic of the system is the double focusing effect revealed by a simple ray diagram. Light transmitted through a slab of thickness $d_2$ located a distance $d_1$ from the source comes to a second focus when,

$$z = d_2 - d_1 \tag{7}$$

The underlying secret of this medium is that both the dielectric function, $\varepsilon$, and the magnetic permeability, $\mu$, happen to be negative. In the instance we have chosen,

$$\varepsilon = -1, \quad \mu = -1 \tag{8}$$

At first sight this simply implies that the refractive index is that of vacuum,

$$n = \sqrt{\varepsilon\mu} \tag{9}$$

but further consideration will reveal that when both $\varepsilon$ and $\mu$ are negative we must choose the negative square root in (9). However the other relevant quantity, the impedance of the medium,

$$Z = \sqrt{\frac{\mu\mu_0}{\varepsilon\varepsilon_0}} \tag{10}$$

retains its positive sign so that when both $\varepsilon = -1$ and $\mu = -1$, the medium is a perfect match to free space and interfaces show no reflection. At the far boundary there is again an impedance match and the light is perfectly transmitted into vacuum.

Calculations confirm that all the energy is perfectly transmitted into the medium but in a strange manner: transport of energy in the +z direction requires that in the medium,

$$k_z' = -\sqrt{\omega^2 c^{-2} - k_x^2 - k_y^2} \tag{11}$$

Overall the transmission coefficient of the medium is,

$$T = tt' = \exp(ik_z'd) = \exp\left(-i\sqrt{\omega^2 c^{-2} - k_x^2 - k_y^2}\,d\right) \tag{12}$$

where *d* is the slab thickness and the negative phase results from the choice of wave vector forced upon us by causality. It is this phase reversal that enables the medium to refocus light by cancelling the phase acquired by light as it moves away from its source.

All this was pointed out by Veselago [1] some time ago. The new message in this letter is that, remarkably, the medium can also cancel the decay of evanescent waves. The challenge here is that such waves decay in *amplitude*, not in *phase*, as they propagate away from the object plane. Therefore to focus them we need to amplify them rather than to correct their phase. We shall show that evanescent waves emerge from the far side of the medium enhanced in amplitude by the transmission process. This does not violate energy conservation because evanescent waves transport no energy, but nevertheless it is a surprising result.

The proof is not difficult. Let us assume S-polarised light in vacuum. The electric field is given by,

$$\mathbf{E}_{0S+} = [0, \ 1, \ 0] \exp(ik_z z + ik_x x - i\omega t) \tag{13}$$

where the wave vector,

$$k_z = +i\sqrt{k_x^2 + k_y^2 - \omega^2 c^{-2}}, \quad \omega^2 c^{-2} < k_x^2 + k_y^2. \tag{14}$$

implies exponential decay. At the interface with the medium some of the light is reflected,

$$\mathbf{E}_{0S-} = r[0, \ 1, \ 0] \exp(-ik_z z + ik_x x - i\omega t) \tag{15}$$

and some transmitted into the medium,

$$\mathbf{E}_{1S+} = t[0, \ 1, \ 0] \exp(ik_z' z + ik_x x - i\omega t) \tag{16}$$

where.

$$k_z' = +i\sqrt{k_x^2 + k_y^2 - \varepsilon\mu\omega^2 c^{-2}}, \quad \varepsilon\mu\omega^2 c^{-2} < k_x^2 + k_y^2. \tag{17}$$

Causality requires that we choose this form of the wave in the medium: it must decay away exponentially from the interface. By matching wave fields at the interface we show that,

$$t = \frac{2\mu k_z}{\mu k_z + k_z'}, \quad r = \frac{\mu k_z - k_z'}{\mu k_z + k_z'} \tag{18}$$

Conversely a wave inside the medium incident on the interface with vacuum experiences transmission and reflection as follows,

$$t' = \frac{2k_z'}{k_z' + \mu k_z}, \quad r' = \frac{k_z' - \mu k_z}{k_z' + \mu k_z} \tag{19}$$

To calculate transmission through both surfaces of the slab we must sum the multiple scattering events,

$$T_S = tt'\exp(ik_z'd) + tt'r'^2\exp(3ik_z'd) + tt'r'^4\exp(5ik_z'd) + \cdots$$

$$= \frac{tt'\exp(ik_z'd)}{1 - r'^2\exp(2ik_z'd)} \tag{20}$$

Substituting from (19) and (20) and taking the limit,

$$\lim_{\substack{\varepsilon \to -1 \\ \mu \to -1}} T_S = \lim_{\substack{\varepsilon \to -1 \\ \mu \to -1}} \frac{tt'\exp(ik_z'd)}{1 - r'^2\exp(2ik_z'd)}$$

$$= \lim_{\substack{\varepsilon \to -1 \\ \mu \to -1}} \frac{2\mu k_z}{\mu k_z + k_z'} \frac{2k_z'}{k_z' + \mu k_z} \frac{\exp(ik_z'd)}{1 - \left(\frac{k_z' - \mu k_z}{k_z' + \mu k_z}\right)^2 \exp(2ik_z'd)} \tag{21}$$

$$= \exp(-ik_z'd) = \exp(-ik_zd)$$

The reflection coefficient is given by,

$$\lim_{\substack{\varepsilon \to -1 \\ \mu \to -1}} R_S = \lim_{\substack{\varepsilon \to -1 \\ \mu \to -1}} r + \frac{tt'r'\exp(2ik_z'd)}{1 - r'^2\exp(2ik_z'd)} = 0 \tag{22}$$

A similar result holds for P-polarised evanescent waves:

$$\lim_{\substack{\varepsilon \to -1 \\ \mu \to -1}} T_P = \lim_{\substack{\varepsilon \to -1 \\ \mu \to -1}} \frac{2\varepsilon k_z}{\varepsilon k_z + k_z'} \frac{2k_z'}{k_z' + \varepsilon k_z} \frac{\exp(ik_z'd)}{1 - \left(\frac{k_z' - \varepsilon k_z}{k_z' + \varepsilon k_z}\right)^2 \exp(2ik_z'd)} \tag{23}$$

$$= \exp(-ik_zd)$$

Thus even though we have meticulously carried through a strictly causal calculation, our final result is that the medium *does* amplify evanescent waves. Hence we conclude that with this new lens *both propagating and evanescent waves contribute to the resolution of the image.* Therefore there is no physical obstacle to perfect reconstruction of the image beyond practical limitations of apertures and perfection of the lens surface. This is the principal conclusion of this letter.

No scheme can be of much interest if the means of realising it are not available. Fortunately several recent developments make such a lens a practical possibility at least in some regions of the spectrum. Some time ago it was shown that wire structures with lattice spacings of the order of a few millimetres behave like a plasma with a resonant frequency, $\omega_{ep}$, in the GHz region [2]. The ideal dielectric response of a plasma is given by,

$$\varepsilon = 1 - \frac{\omega_{ep}^2}{\omega^2} \tag{24}$$

and takes negative values for $\omega < \omega_{ep}$. More recently we have also shown [3] that a structure containing loops of conducting wire has properties mimicking a magnetic plasma,

$$\mu \approx 1 - \frac{\omega_{mp}^2}{\omega^2} \qquad (25)$$

and though the analogy is less perfect, it has been shown that –ve $\mu$ has been attained in these structures [4]. Thus by tuning the design parameters it is certainly possible to produce a structure closely approaching the ideal of,

$$\varepsilon = -1, \quad \mu = -1 \qquad (26)$$

at least at a single frequency.

At optical frequencies several metals behave like a nearly perfect plasma with a dielectric function modelled by (24): silver, gold and copper are perhaps the best examples. The magnetic properties of known materials are less obliging. However we can still make some progress even in this case. Consider the electrostatic limit: a system in which all dimensions are smaller than the wavelength of light. In this system we can neglect radiative effects decoupling electrostatic and magnetostatic fields: electrostatics claims ownership of the P-polarised fields, and magnetostatics of the S-polarised fields.

In the electrostatic limit,

$$\omega << c_0 \sqrt{k_x^2 + k_y^2} = c_0 k_\| \qquad (27)$$

It follows from (14) that,

$$\lim_{k_x^2 + k_y^2 \to \infty} k_z = \lim_{k_x^2 + k_y^2 \to \infty} +i\sqrt{k_x^2 + k_y^2 - \omega^2 c^{-2}}$$
$$= i\sqrt{k_x^2 + k_y^2} \qquad (28)$$

and from (17),

$$\lim_{k_x^2 + k_y^2 \to \infty} k_z' = \lim_{k_x^2 + k_y^2 \to \infty} +i\sqrt{k_x^2 + k_y^2 - \varepsilon\mu\omega^2 c^{-2}}$$
$$= i\sqrt{k_x^2 + k_y^2} = k_z \qquad (29)$$

Hence in this limit we see that for the P-polarise fields dependence on $\mu$ is eliminated and only the dielectric function is relevant. The transmission coefficient of the slab becomes,

$$\lim_{k_x^2 + k_y^2 \to \infty} T_P = \lim_{k_x^2 + k_y^2 \to \infty} \frac{2\varepsilon k_z}{\varepsilon k_z + k_z'} \frac{2k_z'}{k_z' + \varepsilon k_z} \frac{\exp(ik_z' d)}{1 - \left(\frac{k_z' - \varepsilon k_z}{k_z' + \varepsilon k_z}\right)^2 \exp(2ik_z' d)}$$
$$= \frac{4\varepsilon \exp(ik_z d)}{(\varepsilon+1)^2 - (\varepsilon-1)^2 \exp(2ik_z d)} \qquad (30)$$

and hence in this limit we need only assume,

$$\lim_{\substack{k_x^2+k_y^2\to\infty \\ \varepsilon\to-1}} T_P = \lim_{\varepsilon\to-1} \frac{4\varepsilon\exp(ik_zd)}{(\varepsilon+1)^2-(\varepsilon-1)^2\exp(2ik_zd)} = \exp(-ik_zd)$$

$$= \exp\left(+\sqrt{k_x^2+k_y^2}\,d\right)$$

(31)

to obtain focussing of a quasi-electrostatic field, without placing any conditions on $\mu$. It is interesting to note that $\varepsilon = -1$ is exactly the condition for a surface plasmon [5] to exist: there is a link between focussing action and the existence of well defined surface plasmons.

Let us estimate how well we can focus an image using a layer of silver. We shall assume that the object comprises an electrostatic potential with two spikes shown in figure 2. In the absence of the silver the electrostatic potential is blurred at a distance $z = 2d = 80$nm away from the object and we can no longer resolve the two spikes because the higher order Fourier components of the potential are reduced in amplitude,

$$V(x, z = 2d) = \sum_{k_x} v_{k_x} \exp(+ik_xx - 2k_xd) \tag{32}$$

This result is shown in figure 2.

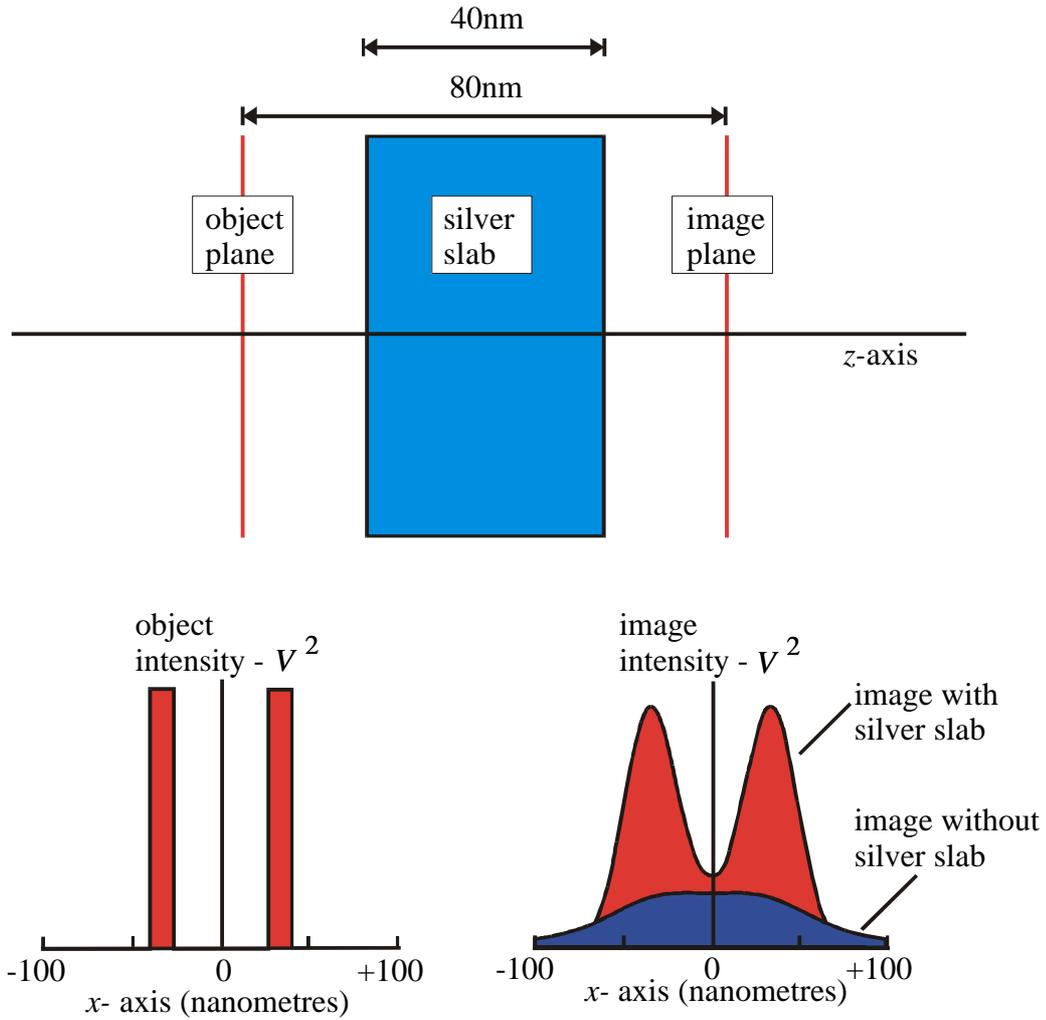

Figure 2. Top: plan view of the new lens in operation. A quasi-electrostatic potential in the object plane is imaged by the action of a silver lens. Bottom: The electrostatic field in the object plane is shown on the left. On the right is the electrostatic field in the image plane with and without the silver slab in place. The reconstruction would be perfect were it not for finite absorption in the silver.

We wish to use a slab of silver, thickness *d*, as a lens to restore the amplitude of the higher order Fourier components and to focus the image. We use the following approximate dielectric function for silver,

$$\varepsilon \approx 5.7 - 9.0^2 \omega^{-2} + 0.4i \qquad (33)$$

Evidently the imaginary part of the dielectric function will place some practical limitations on the focussing effect and making the optimum choice of frequency for focussing,

$$\omega_{sp} = 9.0/\sqrt{6.7} = 3.48 \text{eV} \qquad (34)$$

we have,

$$T_P(k_x) \approx \frac{\exp(-k_x d)}{0.04 + \exp(-2k_x d)} \qquad (35)$$

and the 'focussed' image becomes,

$$V_f(x, z = 2d) = \sum_{k_x} v_{k_x} T_P(k_x) \exp(+ik_x x - 2k_x d)$$
$$= \sum_{k_x} v_{k_x} \frac{\exp(-2k_x d)}{0.04 + \exp(-2k_x d)} \exp(+ik_x x) \quad (36)$$

This result is also plotted in figure 2. Evidently only the finite imaginary part of the dielectric function prevents ideal reconstruction. However considerable focussing is achieved. This result is scale invariant provided that the electrostatic condition (27) is satisfied and provided that the length scale is not so short as to invalidate our assumed form of $\varepsilon$, (31). Doubling or halving all dimensions would give the same result for the image.

Intense focussing of light by exploiting surface plasmons can also be achieved via a completely different route as Ebbesen has recently demonstrated [6,7].

The quasi-static limit also considerably eases design criteria at microwave frequencies. For example we could make a near field electrostatic lens operating in the GHz band using a slab of material containing thin gold wires oriented normal to the surface and spaced in a square lattice cell side 5mm. Perhaps the most interesting possibility for imaging in the GHz band is the magnetostatic limit. A structure comprising a set of metallic rings as described in an earlier paper would give $\mu = -1$ at an appropriate frequency, and would focus sources of magnetic fields into sharp images. Since many materials are transparent to magnetic fields this would make an interesting imaging device for peering inside non-magnetic objects.

We have given a prescription for bringing light to a perfect focus without the usual constraints imposed by wavelength. This is achieved by recognising that the recently discovered negative refractive index material restores not only the phase of propagating waves but also the amplitude of evanescent states. For very short distances the electrostatic or magnetostatic limits apply enabling a practical implementation to be simulated in the form of a slab of silver. This device focuses light tuned to the surface plasma frequency of silver and is limited only by the resistive losses in the metal. We do not doubt that there are many further practical consequences of this concept.

I thank David Smith, Sheldon Schultz and Mike Wiltshire for valuable correspondence on the concept of negative refractive index.